\documentstyle[12pt]{article}
 
\def\be{\begin{equation}}
\def\ee{\end{equation}}
\def\a{\alpha}
\def\s{\sigma}

\def\c{\tilde{c}}
\def\b{\tilde{b}}
\def\n{\{n\}}
 
\def\O{\hat{O}}
\def\Ot{\hat{\tilde{O}}} 
\def\F{\hat{F}}
\def\l{\lambda} 
\def\vac{|0\rangle} 
\def\aa{{\large a}}
\def\ra{\rangle}
\def\la{\langle}

\begin{document}

\begin{center} 
{\large Direct proof for the Scalar Product with Bethe \\
              eigenstate in Spin chains. }
\end{center}

\begin{center}
{\large  A.A. Ovchinnikov  }
\footnote{E - mail address: ovch@ms2.inr.ac.ru}
\end{center}

\begin{center}
{\it Institute for Nuclear Research, RAS, Moscow.}
\end{center}

\begin{abstract}

We present the simple and direct proof of the determinantal 
formula for the scalar product of Bethe eigenstate with an 
arbitrary dual state. We briefly review the direct calculation 
of the general scalar product with the help of the factorizing 
operator and the construction of the factorizing operator itself. 
We also comment on the previous determination of the scalar product of 
Bethe eigenstate with an arbitrary dual state.   

\end{abstract}

\vspace{0.2in} 

              {\bf 1. Introduction.}  

\vspace{0.2in}

One of the most important open problems in the theory of quantum 
integrable models is the calculation of the correlation functions. 
In the framework of the Algebraic Bethe Ansatz method \cite{FST} 
the problem is the combinatorial complexity of calculations due to 
the structure of Bethe eigenstates. In spite of these difficulties 
many important results have been obtained. Among them one 
can mention the derivation by Korepin the Gaudin formula for the norm of Bethe 
eigenstates \cite{K1} and the calculation of the scalar product of Bethe 
eigenstate with an arbitrary dual state \cite{S}, which leads straightforwardly 
to the determinant representation of the formfactors of local operators  
(see \cite{KMT} for the case of spin-1/2 chains and \cite{Kojima}, \cite{KS}  
for the case of Bose-gas with $\delta$ - function interaction).     
Using the Algebraic Bethe Ansatz and the solution of the quantum inverse
scattering problem the authors of ref.\cite{KMT1} obtained the multiple
integral representation for the correlation functions found previously 
using the other methods (see for example \cite{M} and references therein).
The concept of factorizing F - matrix introduced recently by 
Maillet and Sanches de Santos \cite{MS} following the concept of Drinfeld's 
twists in his theory of Quantum Groups and construction of the Monodromy 
matrix in the F -basis \cite{MS}, \cite{KMT}, \cite{O} allows in some respect 
to simplify the calculations in comparison with the general theory 
of scalar products developed previously \cite{K1}, \cite{IK}. 

  At the same time only the very limited number of physical results 
for the correlation functions (or formfactors) have been obtained from 
the first principles. So the developing of new calculational methods 
and better understanding of the underlying mathematical structure for the 
expressions for the correlators is required. For example it is still 
not completely clear if it is possible to obtain the determinantal expressions 
for the correlators without using the auxiliary dual fields \cite{K1}, \cite{IK}. 

  The aim of the present letter is to present the simple and direct proof 
for the expression of the scalar product of Bethe eigenstate with an 
arbitrary dual state. First, to fix the notations and derive the 
expression for an arbitrary scalar product, we briefly review the results 
of ref.\cite{O} on the construction of the factorizing operator (Section 2), 
and simple calculation of the general scalar product (Section 3). Next, we 
present the new version of the proof given by Slavnov \cite{S} for the scalar 
with Bethe eigenstate. Finally in Section 4 we present the direct and a simple 
proof of this formula. Our results could shed some light on the possibility 
to obtain the determinantal expressions for various correlators.

\vspace{0.2in}

     {\bf 2. Construction of the factorizing operator.}  

\vspace{0.2in}

    We consider in this letter the XXX or XXZ spin- 1/2 chains of finite 
length N.  In the present section we diagonalize the operator $A$ and introduce 
the factorizing operator.  Let us fix the notations: the normalization 
of basic S - matrix, the definition of monodromy matrix and write down the 
Bethe Ansatz equations. For the rational case (XXX- chain) the S- matrix has the 
form $S_{12}(t_1,t_2)=t_1-t_2+\eta P_{12}$, where $P_{12}$ is the permutation 
operator. In general (XXZ) case it can be written as 
\[
S_{12}(t_1,t_2) =  \left( 
\begin{array}{cccc}  
a(t) & 0 & 0 & 0 \\
0 & c(t) & b(t) & 0 \\
0 & b(t) & c(t) & 0 \\
0 & 0 & 0 & a(t) 
\end{array}  \right)_{(12)}, ~~~~ t=t_1-t_2. 
\]
One can choose the normalization $a(t)=1$ so that the functions $b(t)$ and $c(t)$ 
become $\b(t)=\phi(\eta)/\phi(t+\eta)$, $\c(t)=\phi(t)/\phi(t+\eta)$, 
where $\phi(t)=t$ for the isotropic (XXX) chain and $\phi(t)=\sinh(t)$ 
for the XXZ- chain. With this normalization the $S$-matrix satisfies 
the unitarity condition $S_{12}(t_1,t_2)S_{21}(t_2,t_1)=1$. 
The monodromy matrix is defined as 
\[
T_{0}(t,\{\xi\})= S_{10}(\xi_1,t)S_{20}(\xi_2,t)...S_{N0}(\xi_N,t), 
\]
where $\xi_i$ are the inhomogeneity parameters. We define the operator entries 
in the auxiliary space $(0)$ as follows: 
\[
\la \beta |T_0|\a \ra = \left( 
\begin{array}{cc}
A(t) & B(t) \\
C(t) & D(t) 
\end{array} \right)_{\a\beta}; ~~~~\a,\beta = (1,2) = (\uparrow; \downarrow). 
\]
We denote throughout the paper $(\uparrow;\downarrow)=(1;0)$ so that the 
pseudovacuum (quantum reference state) $\vac=|\{00...0\}_N\ra$. 
The triangle relation (Yang-Baxter equation) reads:  
\[
S_{12}S_{13}S_{23}=S_{23}S_{13}S_{12},~~~R_{00^{\prime}}T_0 T_{0^{\prime}}= 
T_{0^{\prime}}T_0 R_{00^{\prime}}; ~~~
R_{00^{\prime}}=P_{00^{\prime}}S_{00^{\prime}}. 
\]
The action of the operators on the pseudovacuum is: $A(t)\vac=\aa(t)\vac$ 
($\aa(t)=\prod_{\a}\c(\xi_{\a}-t)$), $D(t)\vac=\vac$, $C(t)\vac=0$. 
The Bethe Ansatz equaions for the eigenstate of the Hamiltonian
$\prod_{i=1}^M B(t_i)\vac$ and the corresponding eigenvalue of the 
transfer - matrix $Z(t)=A(t)+D(t)$ are 
\[
\aa(t_i)=\prod_{\a\neq i} \c(t_{\a}-t_i) (\c(t_i-t_{\a}))^{-1},~~~~~ 
\Lambda(t,\{t_{\a}\})=\aa(t)\prod_{\a=1}^{M}\c^{-1}(t_{\a}-t)+ 
\prod_{\a=1}^{M}\c^{-1}(t-t_{\a}),   
\]
where $t_{\a}$ are the solution of Bethe Ansatz equations. 

 In order to construct the operator $\O = \O_{1...N}$ which diagonalizes the 
operator $A(t)$ ($\O^{-1} A \O =diag(A)$), let us first construct the 
eigenfunctions of the operator $A$. One can do it in two different ways. 
First, note that $A$ - is a triangular matrix in a sense that it makes particles 
(the spin-up - coordinates) move to the right on the lattice $1...N$. 
Thus its eigenvalues are coincide with its diagonal matrix elements and 
therefore are charactarized by the set of integers ($n_1,...n_M$) - the spin-up 
positions. Let us denote this eigenfunctions by $|\phi (n_1,...n_M)\ra$. Clearly, 
$|\{00..0\}_{N-M}\{11..1\}_{M}\ra$ is an eigenstate of $A(t)$. 
Therefore, considering the permutation 
\[
S_{10}...S_{N0} \rightarrow  
S_{10}...S_{N0}S_{n_M 0}...S_{n_1 0} ~~~~(n_1 < n_2 <...< n_M),
\]
we realize that 
\be
|\phi (n_1,...n_M)\ra = T_{n_1}T_{n_2}...T_{n_M} |n_1,n_2,...n_M\ra 
\label{tn}
\ee
where 
\[
T_n = S_{n+1,n}S_{n+2,n}...S_{Nn}, 
\]
is an eigenstate of the operator $A(t)$. Note that if we modify the given 
permutation interchanging $n_i\leftrightarrow n_j$ the operator in the right-hand 
side of (\ref{tn}) modifies but the state $|\phi(\n)\ra$ remains the same. For example,
for $M=2$: $T_{n_2}T_{n_1}^{\prime}=T_{n_1}T_{n_2}S_{n_{2}n_{1}}^{-1}$, where $n_1<n_2$ 
and the prime means the absence of the term $S_{n_{1}n_{2}}$ in $T_{n_1}$. Since 
$S_{n_{1}n_{2}}|n_1,n_2\ra=|n_1,n_2\ra$ the state remains the same.      

The second way - is to consider the state 
\be
|\phi (n_1,...n_M)\ra = B(\xi_{n_1})B(\xi_{n_2})...B(\xi_{n_M}) \vac,~~~(n_i \neq n_j). 
\label{bb}
\ee
Using the fundamental commutation relation:
\be
A(t)B(q)= \frac{1}{\c(q-t)}B(q)A(t) - \frac{\b(q-t)}{\c(q-t)}B(t)A(q),
\label{ab}
\ee
and the fact that for the pseudovacuum state $A(\xi_i)\vac =0$, we find again 
that $|\phi(\n)\ra$ ($\n=\{n_1,...n_M\}$) is an eigenstate of $A(t)$ with the 
eigenvalue $A_{\n\n}(t)=\prod_{\a\neq n_k}\c(\xi_{\a}-t)$.  

One can see that the states $|\phi(\{n\})\ra$ (\ref{tn}) and (\ref{bb}) are coincide. 
This can be seen using the identity 
\be
B(\xi_n)=T_n \la 0|S_{10}\ldots S_{n-1,0}P_{n0}|1\ra 
\label{bxi}
\ee
and ordering the (commuting) operators $B(\xi_{n_i})$ in eq.(\ref{bb}) 
according to the prescription $n_1<n_2<...<n_M$, so that the second operator 
in the last formula simply creates the particle (spin- up) at the site $n$ 
with the amplitude equal to unity.   

  Now we can introduce the diagonalizing operator $\O=\O_{1...N}$ (we  
show later that it is also the factorizing operator). Let us define the 
operator $\O$ such that 
\[
|\phi(\{n\})\ra = \O(\{n\})|\{n\}\ra = \O |\n\ra , 
\]
where $\O(\{n\})$ is given by the equation (\ref{tn}). Clearly the operator 
$A^{F}(t)=\O^{-1}A(t)\O$ - is diagonal. Now we immediately get 
the simple formula for the matrix elements of the factorizing operator: 
\be
\O_{\{m\}\{n\}} = \la\{m\}|B(\xi_{n_1})B(\xi_{n_2})...B(\xi_{n_M}) \vac, 
\label{me}
\ee 
where $\{m\}=\{m_1,...m_M\}$. From this expression (see (\ref{bxi})) one 
see that $\O$ is the triangular matrix (in the same sense as the operator 
$A((t)$). 
Let us show that the operator $\O$ is invertible and construct the inverse 
operator. To find this operator we define the operator $\Ot$ analogously to 
the previous case as $\la\tilde{\phi}(\{n\})|=\la\{n\}|\Ot$ where the 
corresponding dual states and the matrix elements of the operator $\Ot$ are: 
\be
\la\tilde{\phi}(n_1,...n_M)|=\la 0|C(\xi_{n_1})...C(\xi_{n_M}),~~~~ 
\Ot_{\{m\}\{n\}}=\la 0|C(\xi_{m_1})C(\xi_{m_2})...C(\xi_{m_M})|\n\ra. 
\label{mec}
\ee  
Let us calculate the following scalar product  
\be
\la\tilde{\phi}(\{m\})|\phi(\{n\})\ra = \la \{m\}|\Ot\O|\{n\}\ra =
\la 0|C(\xi_{m_1})...C(\xi_{m_M})
B(\xi_{n_1})...B(\xi_{n_M})\vac. 
\label{oo}
\ee
This scalar product can be calculated using another well known 
relation for the commutator $\left[ B(q),C(t)\right]$ following from 
the Yang-Baxter equation, and using again that for any site $i$, 
$A(\xi_i)\vac=0$. Moving  the operators $A$ and $D$ to the right, 
repeating consequently the relation (\ref{ab}) and the analogous relation for 
$D$ (which differs from eq.(\ref{ab}) by the interchange $t\leftrightarrow q$ 
in the coefficients between the products of the operators)  
and using the equations 
$C(\xi_n)B(\xi_n)\vac =\prod_{\a\neq n}\c(\xi_{\a}-\xi_{n})\vac$, 
we obtain that the matrix $\hat{f}=\Ot\O$ is diagonal.  
The corresponding diagonal matrix elements can be found either using the 
procedure mentioned above or, which is the simplest way, using the 
representation (\ref{bxi}) (and the same for the operator $C(t)$):  
\be
f(n_1,...n_M)=\prod_k\left(\prod_{\a\neq n_k,n_j}\c(\xi_{\a}-\xi_{n_k})\right) . 
\label{fme}
\ee
Thus we obtained the inverse matrix $\O^{-1}$:
\[
\Ot\O=\hat{f},~~~~~ \O^{-1}=\hat{f}^{-1}\Ot. 
\]
Before proceeding with evaluation of the matrix elements of the other operators 
in the new  basis, let us mention some usefull properties of the operator $\O$, 
and prove that, in fact, it is the factorizing operator in a sense of 
the definition \cite{MS}.      
First, $\O$ and $\O^{-1}$ are the triangular matrices (upper triangular as $A(t)$). 
Second, the pseudovacuum state is an eigenstate of $\O$ ($\O^{-1}$) with the 
eigenvalue equal to unity. In general we have the following equations for 
arbitrary number of particles $n$:
$\O|\{00..0\}_{N-n}\{11..1\}_n\ra =|\{00..0\}_{N-n}\{11..1\}_n\ra$, 
the similar formulas for the operator $\Ot$ and the same formulas for the inverse 
operators. From this formulas one can already suspect that $\O$ is the factorizing 
operator. Indeed for the particular permutation (\ref{tn}) $\sigma(\n)$
the factorizing condition is represented as 
$\O(\O^{\sigma(\n)})^{-1}=\O(\n)$, where $\O(\n)=T_{n_1}..T_{n_M}$ and is fulfilled 
at least for the state $|\n\ra$ since due to the last formulas 
$\O^{\sigma(\n)})^{-1}|\n\ra=|\n\ra$. The rigorous proof goes as follows. We  construct
the operator that acting on the state $|\n\ra$ produces the state $\O(\n)|\n\ra$. 
It is easy to see that the operator that fulfills the above requirement is: 
\be
\O_{12\ldots N}=\F_1\F_2\ldots \F_N,~~~~~\F_i = (1-\hat{n}_i)+T_i \hat{n}_i,  
\label{f}
\ee
where $\hat{n}_i$ is the operator of the number of particles (spin up) at the 
given site $i$. The operators $\F_i$ entering (\ref{f}) do not commute 
and their ordering in eq.(\ref{f}) is important. To prove the factorizing property 
of this operator it is sufficient to consider only one particular permutation, 
say the the permutation $(i,i+1)$ since all the others can be obtained as a 
superposition of these ones for different $i$. One can show \cite{MS}, \cite{O} that  
\be
\O=S_{i+1,i}\O^{(i,i+1)}, 
\label{ii}
\ee
Evidently, for any transmutation $\sigma\in S_N$  we will obtain only one 
operator $R^{\s}_{1...N}$ (the operator constructed from $S$- matrices)  
on the left of the operator $\O$. Thus $\O$ is the factorizing operator 
in a sense of \cite{MS}.  

   Let us  calculate the matrix elements of $B(t)$ and $C(t)$ - operators 
in the F - basis: $B^{F}(t)=FB(t)F^{-1}=\O^{-1}B(t)\O$ (and the same for $C(t)$). 
The general scheme to perform the calculations is to use the formalism developed  
previously, which leads to the following equation: 
\be
B(t) B(\xi_{n_1})...B(\xi_{n_M}) \vac = \sum_{x} \left( 
B(\xi_{x})B(\xi_{n_1})...B(\xi_{n_M}) \vac \right) \phi(x,t,\n), 
\label{def}
\ee
where $\phi(x,t,\n)$- is exactly the matrix element of the operator 
$B(t)$ in the new basis. To get the single term in the sum in 
eq.(\ref{def}), we act by the operator $A(\xi_x)$ ($x\neq n_k$) 
at both sides of this equation. Using again the 
property $A(\xi_i)|0\ra=0$ and eq.(\ref{ab}) for the left-hand side 
of (\ref{def}) we get the single term with $B(\xi_x)$ for the right-hand side,  
which can be evaluated using eq.(\ref{ab}) and the formula 
$A(\xi_x)B(\xi_x)|0\ra=\prod_{\a\neq x}\c(\xi_{\a}-\xi_x)B(\xi_x)|0\ra$ , 
which can be proved by direct computations. After the cancellation of similar 
terms at both sides of eq.(\ref{def}) we get the matrix element $\phi(x,t,\n)$ 
and finally obtain in the operator form: 
\be
B^{F}(t)=\sum_{x}\sigma_{x}^{\dagger}~\b(\xi_{x}-t)\prod_{\a\neq x} 
\left( \begin{array}{ll}
\c(\xi_{\a}-t)(\c(\xi_{\a}-\xi_{x}))^{-1},     & \a \neq n_k \\
    1,                                          & \a = n_k 
\end{array} \right).  
\label{bf}
\ee
With this expression one can prove the equation 
$\prod_i B^{F}(\xi_{n_i})\vac=|\n\ra$ which is consistent with  
the formulas of the previous section. 
For the operator $C(t)$ proceeding in a similar way and using the 
relation for the commutator $\left[B;C\right]$, we get after some algebra 
\be
C^{F}(t)=\sum_{x}\sigma_{x}^{-}~\b(\xi_{x}-t)\prod_{\a\neq x} 
\left( \begin{array}{ll}
\c(\xi_{\a}-t),                   & \a \neq n_k \\
(\c(\xi_x-\xi_{\a}))^{-1},     & \a = n_k 
\end{array} \right).  
\label{cf}
\ee
The operators (\ref{bf}) and (\ref{cf}) are quasilocal i.e. they describe the 
flipping of the spin on a single site with the amplitude depending on the 
positions of spins on all the other sites of the chain. The operator 
$D^F(t)$ can be found using either the same method or the quantum determinant 
relation and has a (quasi)bilocal form. 

\vspace{0.2in}

           {\bf 3. Calculation of the scalar products.}

\vspace{0.2in}

  In this section we use the developed formalism to obtain the expressions for  
the correlation functions for the spin chains. We use the factorizing operator 
and, in particular, the expression for its matrix elements (\ref{me}) to obtain 
the expression for the general correlation function (scalar product) 
\[
S_M(\{\l\},\{t\})=\la 0|C(\l_1)C(\l_2)...C(\l_M)
                    B(t_1)B(t_2)...B(t_M)\vac ,  
\] 
where $\{\l\}$ and $\{t\}$ are two arbitrary sets of parameters (not necessarily 
satisfying the Bethe Ansatz equations). The correlation function can be 
represented in the following form 
\[
\la 0|Tr_{0_1,...0_{2M}}\left(\sigma_{0_1}^{+}...\sigma_{0_M}^{+} 
\sigma_{0_{M+1}}^{-}...\sigma_{0_{2M}}^{-} 
T_{0_1}\ldots T_{0_{2M}} \right)  \vac 
\]
The auxiliary spaces $0_1,\ldots 0_{2M}$ can be considered as a lattice 
consisting of $2M$ sites with the corresponding spectral parameters 
$\l_1,\ldots\l_M,t_1,,\ldots t_M$. Rearranging the basic $S$- matrices 
entering the product of the monodromy matrices we arrive at the operator 
\[
\tilde{T}_1 \ldots \tilde{T}_N,
\]
where the new monodromy matrices act in the auxialiary space instead of the 
original quantum space:
\[
\tilde{T}_n = S_{n0_1}S_{n0_2} \ldots S_{n0_{2M}}.
\]
Obviously using this matrices the correlator can be 
represented as the following matrix element 
in the new quantum space $0_1,...0_{2M}$:  
\[
\la \{00..0\}_M\{11..1\}_M|\tilde{A}(\xi_1)\ldots\tilde{A}(\xi_N) 
|\{11..1\}_M\{00..0\}_M \ra  
\]
(we use the symmetry $A(t)\leftrightarrow D(t)$, $0\leftrightarrow 1$ here). 
Transforming the operators $\tilde{A}(\xi_i)$ to the F - basis we find 
at $\xi_i=0$: 
\[ 
S_M = \sum_{\n} \la \{00..0\}_M\{11..1\}_M|\O|\n\ra
\la\n| \left(\tilde{A}^{F}(0))\right)^N |\n\ra
\la\n| \O^{-1} |\{11..1\}_M\{00..0\}_M \ra . 
\]
The sum is over the states labeled by the positions of M particles  
$\n=n_1\ldots n_M$ on a lattice consisting of $2M$ sites with the 
inhomogeneity parameters $\l_1\ldots\l_M,t_1\ldots t_M$. We use  
$\O^{-1}=\hat{f}^{-1}\Ot$ and the representations (\ref{me}), (\ref{mec}) 
for the matrix elements in the last formula. Let us denote by 
$\mu_1,\ldots\mu_M$ the parameters (from the set $(\{\l\},\{t\})$) 
corresponding to the sites $n_1\ldots n_M$, and by $\nu_1,\ldots\nu_M$ 
the rest of the parameters so that $\{\l\}\cup\{t\}=\{\mu\}\cup\{\nu\}$. 
The first matrix element in the formula for
$S_M$ is equal to 
\be
\la \{00..0\}_M\{11..1\}_M| B(\mu_1)\ldots B(\mu_M) \vac =
\la\{11..1\}_M| B^{\prime}(\mu_1)\ldots B^{\prime}(\mu_M) \vac, 
\label{z}
\ee
where $B^{\prime}(\mu)$ are the same operators difined on the lattice 
consisting of $M$ sites with the inhomogeneity parameters $t_1\ldots t_M$. 
The second matrix element can be reduced to the same expression with the 
parameters $\l_1\ldots \l_M$ (using the symmetry $C\leftrightarrow B$, 
$0\leftrightarrow 1$). Then using the formula for the matrix elements of the 
operator $\hat{f}^{-1}$ (see eq.(\ref{fme})) we finally obtain the 
expression: 
\be
S_M(\{\l\},\{t\})=\sum_{n_1,..n_M}\left(\prod_j a(\nu_j)\right)  
\Phi_{M}(t,\mu)\Phi_{M}(\l,\mu)
\prod_{i,j} \frac{1}{\c(\mu_i-\nu_j)}, 
\label{final}
\ee
where we denoted by $\Phi_{M}(\xi,t)$ the functions in the right hand side of 
eq.(\ref{z}). The determinant representation of this function and its 
properties are well known (see for example \cite{K1},\cite{Coker}
\cite{KMT},\cite{O}): 
\be
\Phi_{M}(\xi,t)=\frac{\prod_{i,j}(t_i-\xi_j)}{\prod_{i<j}(t_i-t_j)
\prod_{j<i}(\xi_i-\xi_j)}\det_{ij}\left(\frac{\eta}{(t_i-\xi_j)
(t_i-\xi_j+\eta)}\right)
\label{satur}
\ee
for the rational case. Here $\{\xi\}$ are the inhomogeneity parameters 
and $\{t\}$ are the arguments of $B$- operators (see eq.(\ref{z})). Note 
that in the expression (\ref{final}) all the dependence on the inhomogeneity 
parameters $\xi_i$ is contained only in the functions $a(\nu_i)$. 
The functions $a(\nu)$ in eq.(\ref{final}) are exactly the functions 
defined above $a(\nu)=\prod_{\a}\c(\xi_{\a}-\nu)$ while in the rest 
of this formula due to the definition of the matrices $\tilde{T}_i$ 
one should interchange the arguments in the functions $\c^{-1}(\nu_i-\mu_j)$ 
which is taken into account in eq.(\ref{final}) 
(or one could make the replacement $\eta\rightarrow-\eta$).  
Using the properties of the functions $\Phi_M$ one can represent the 
general formula (\ref{final}) in a different way:  
\[
\sum_{m=0}^{M}\sum_{k,n}\left(\prod \left( a(\l_n)a(t_k)\right)\right)
\Phi_{m}(t_k,\l_{\beta})\Phi_{M-m}(\l_{n},t_{\a})
\prod \frac{1}{\c(\l_{\beta}-\l_n)} \frac{1}{\c(t_{\a}-t_k)}
\frac{1}{\c(t_{\a}-\l_n)} \frac{1}{\c(\l_{\beta}-t_k)},  
\]
where the sum is over the two sets $k_1,...k_m$ and $n_1,...n_{M-m}$. 
We used the following simplified notations in this formula. We devided 
the set $\{t\}$ into two subsets $\{t\}=\{t_k\}\cup\{t_{\a}\}$ where 
$\{t_k\}=(t_{k_1}...t_{k_m})\in\{\nu\}$,  
$\{t_{\a}\}\in\{\mu\}$  and analogously $\{\l\}=\{\l_n\}\cup\{\l_{\beta}\}$, 
$\{\l_n\}=(\l_{n_1}...\l_{n_{M-m}})\in\{\nu\}$, $\{\l_{\beta}\}\in\{\mu\}$.  
The products in the last formula are over the indices labeling the elements 
of the corresponing sets. One can further rewrite this formula using 
the expressions for the functions $\Phi_m$ to get the formula which leads 
to the determinant representation for $S_M$ after the special dual fields 
are introduced \cite{IK}. 

Let us derive the scalar product of the Bethe eigenstate 
with an arbitrary dual state along the lines of ref.\cite{S} starting from 
the formula (\ref{final}). The set of the parameters $\{t\}$ obey the 
Bethe equations  so for each $a(t_i)$ in the sum (\ref{final}) one can 
substitute the function 
\[
f(t_i)=\prod_{\a\neq i}\frac{\c(t_{\a}-t_i)}{\c(t_i-t_{\a})}=
\prod_{\a\neq i}\frac{\phi(t_{\a}-t_i-\eta)}{\phi(t_{\a}-t_i+\eta)}. 
\]
Then the idea is that one can calculate the sum in (\ref{final}) for an 
arbitrary smooth function $a(\l)$, which behaves at least as a constant 
at infinity, used for the terms $a(\l_j)$ 
(not necessarily equal to $a(\l)=\prod_{\a}\c(\xi_{\a}-\l)$). 
In that case the sum has the simple poles in the variables 
$\{t\}$ and $\{\l\}$ at the points $t_i=\l_j$ and don't have any other poles,  
for example the poles at $\l_i=\l_j$, which exist if, according to \cite{S} 
one considers $a(\l_j)$ as an independent variables. These poles was  
not considered in \cite{S} (actually only the behaviour of the function 
in the parameters $t_i$ was considered - obviously there is no poles at 
$t_i=t_j$ if the Bethe Ansatz equations are taken into account). 
One can use the symmetry of $S_M$ in $\{t\}$ and $\{\l\}$ and single out the 
residue at $\l_1\rightarrow t_1$ (which is contained in the term with 
$\l_1\in\{\nu\}$, $t_1\in\{\mu\}$ and vice versa):  
\[
S_M \left( \{\l\},\{t\},a(\nu_j) \right)  \vert_{\l_1\rightarrow t_1}
\rightarrow \eta \frac{a(\l_1)-f(t_1)}{t_1-\l_1} 
\prod_{\a\neq 1} \frac{1}{\c(t_{\a}-t_1)} \frac{1}{\c(\l_{\a}-t_1)}
S_{M-1} \left( \{\l\}',\{t\}',a'(\nu_j) \right), 
\]
where $\{\l\}'$, $\{t\}'$ do not contain  $\l_1$, $t_1$ and 
\[
a'(\nu_j)=a(\nu_j) \frac{\c(\nu_j-t_1)}{\c(t_1-\nu_j)}. 
\]
The functions $f(t_i)$ entering the sum $S_{M-1}$ are also modified 
so that the terms with $t_1$ are absent in their definition. 
Also the behaviour of $S_M\sim 1/\l_1$ at $\l_1$ going to infinity 
should be taken into account.   
It can be easily proved that the same recurrence relation at  
$\l_1\rightarrow t_1$ is obeyed by the following expression: 
\[
S_M(\{\l\},\{t\})=\frac{1}{\prod_{i<j}(t_i-t_j)\prod_{j<i}(\l_i-\l_j)}
\det_{ij}(M_{ij}(t,\l))
\]
\be
M_{ij}(t,\l) = \frac{\eta}{(t_i-\l_j)} 
 \left( a(\l_j)\prod_{\a\neq i}(t_{\a}-\l_j +\eta) - 
 \prod_{\a\neq i}(t_{\a}-\l_j - \eta) \right),  
\label{mbethe} 
\ee
where at the end of calculations one should take the actual 
form of the function $a(\l)$. So we proved that for an arbitrary 
function $a(\l)$ the residues of the only existing poles at 
$t_i=\l_j$ satisfy the the same recursion relation in both 
formulas. Since the function (in each variable) is completely 
determined by the positions of (all) poles and the corresponding 
residues (along with the behaviour at the infinity) the two 
functions (\ref{mbethe}) and (\ref{final}) should coincide for 
an arbitrary function $a(\l)$.   
To complete the proof, one should check the formula (\ref{mbethe}) 
for $M=1,2$. Equivalently, one can check the coefficients 
corresponding to the terms $\prod_{i=1}^{M}a(\l_i)$ and 
$\prod_{i=1}^{M}a(t_i)=1$ in eq.(\ref{mbethe}) which are 
in agreement with (\ref{final}). 
The formula (\ref{mbethe}) is written for the rational case. 
This formula can also be represented through the Jacobian 
$\det_{ij}(\partial/\partial t_i\Lambda(\l_j;\{t_{\a}\})$. 
The orthogonality of two different Bethe eigenstates was shown 
in ref.(\cite{KS}). From the equation (\ref{mbethe}) taking 
the limit $\l_i\to t_i$ one can easily obtain 
the formula for the norm of the Bethe eigenstate \cite{K1}.

\vspace{0.7in}

{\bf 4. Direct proof for the scalar product with Bethe eigenstate.}

\vspace{0.2in}

Here we present the direct and the simple proof of the determinant expression 
(\ref{mbethe}) for the scalar product with one Bethe eigenstate. 
Although the proof given in the previous section is correct the generalization 
of this method to the expressions for different correlation functions is 
not straightforward. It is still an open question if it is possible to 
obtain the determinant expressions for various correlation functions 
(the simlest example is the emptiness formation probability, which can be 
represented as a sum very similar to that of eq.(\ref{final})) without the 
auxiliary dual fields \cite{K1},\cite{IK}. Thus the various direct derivations 
of the formula (\ref{mbethe}) are worth of developing. 
Let us present the direct proof of this formula based on the 
expression (\ref{final}). 

First, let us rewrite the expression (\ref{final}) for the case when 
the parameters $\{t\}$ satisfy the Bethe Ansatz equations.  
Expressing the functions $a(t_k)=f(t_k)$ and using 
the explicit form of the function $\Phi_M$ we get: 
\[
S_M(\l,t)=\frac{1}{\prod_{i<j}(t_i-t_j)}\frac{1}{\prod_{j<i}(\l_i-\l_j)}       
\sum_{k,n} (-1)^{P_k}(-1)^{P_n} \prod(a(\l_n)) 
\]
\[
\det(\l_{\beta};t_k) \det(t_{\a};\l_n)
\prod(\l_{\beta}-\l_n+\eta)(t_k-t_{\a}+\eta)(t_{\a}-\l_n+\eta)
(\l_{\beta}-t_k+\eta), 
\]
where again the products is over all elements of the corresponding 
sets of parameters. Here the notations are exactly the same as in 
the previous section. The sign factors $P_k$, $P_n$ depend on the 
sets of the coordinates $k_1,\ldots k_m$, $n_1,\ldots n_{M-m}$ in 
order to obtain the factors in front of the sum from the terms 
$\prod(t_k-t_{\a})^{-1}$ and $\prod(\l_{\beta}-\l_n)^{-1}$ 
We also used the simplified notations for the determinants:  
\be
\det(t,\l)= \det_{ij}\left(\frac{\eta}{(t_i-\l_j)(t_i-\l_j+\eta)}\right).
\label{det}
\ee
The order of lines and columns in the determinants 
$\det(\l_{\beta};t_k)$, $\det(t_{\a};\l_n)$ corresponds to the prescription
$k_1<k_2<\ldots <k_m$ and the same for the other sets of variables. 
Another, equivalent way to rewrite the last formula for $S_M$ is:  
\be
S_M(\l,t)=\frac{1}{\prod_{i<j}(t_i-t_j)}\frac{1}{\prod_{j<i}(\l_i-\l_j)}       
\sum_{k,n} (-1)^{P_k}(-1)^{P_n} (-1)^{Mm} \prod(a(\l_n)) 
\label{dd}
\ee
\[
\prod (t-\l_n+\eta) \prod (t-\l_{\beta}-\eta)
\det(\l_{\beta};t_k) \det(t_{\a};\l_n)
\prod \left( \frac{t_k-t_{\a}+\eta}{t_k-\l_n+\eta}  \right)
\prod \left(\frac{\l_{\beta}-\l_n+\eta}{\l_{\beta}-t_{\a}+\eta} \right),   
\]
where again the products is over all elements of the corresponding 
sets of parameters. Here again the sign factors $P_k$, $P_n$ depend on the 
sets of the coordinates $k_1,\ldots k_m$, $n_1,\ldots n_{M-m}$  and 
we also used the simplified notations for the determinants (\ref{satur})   
and denote for example 
$\prod(t-\l_n+\eta)=\prod_j\prod_{i=1}^{M}(t_i-\l_{n_{j}}+\eta)$. 
In this formula the poles at $t_i=\l_j$ are contained in the determinants. 
If the two last products are dropped out this formula becomes the result of 
the decomposition of the determinant in eq.(\ref{mbethe}). Then  
one can see explicitly that the residues in the poles in the formulas 
(\ref{final}) and (\ref{mbethe}) satisfy the same recurrence relations. 

Our proof of the formula (\ref{mbethe}) is based on the on the 
expression (\ref{dd}). The idea is to calculate the sum over the set 
of the coordinates $\{k\}$.  
For example at the first stage consider the case $m=M-1$ which corresponds 
to the terms containing the single function $a(\l_{n_1})$ 
(the set $\{\l_n\}$ contains only one element $\l_{n_1}$ 
and the set $\{t_{\a}\}$ - only one element $t_{\a_1}$). 
Then the sum over the set $\{k\}$ in eq.(\ref{dd}) has the following form: 
\be
a(\l_{n_1})\sum_{\a_1} \det(\l_{\beta};t_k)
\frac{(-1)^{\a_1}\eta}{(t_{\a_1}-\l_{n_1})(t_{\a_1}-\l_{n_1}+\eta)}
\prod\left(\frac{t_k-t_{\a_1}+\eta}{t_k-\l_{n_1}+\eta}\right)
\prod\left(\frac{\l_{\beta}-\l_{n_1}+\eta}{\l_{\beta}-t_{\a_1}+\eta}\right). 
\label{m}
\ee
This expression can be considered as a first column developement of 
the determinant of some new $M\times M$ matrix, the parameter 
$\l_{n_1}$ being assigned to the first column of this matrix. 

To find this matrix let us first  consider the determinant 
\[
\det_{ij}(M_{ij})=\det_{ij}\left(\frac{1}{(t_i-\l_j)(t_i-\l_j+\eta)}\right),  
\]
and modify the first column of the matrix $M_{i1}$ adding 
a linear combination of the other columns so that the determinant 
$\det M$ is unchanged: 
\[
M'_{i1}=M_{i1}+\sum_{x\neq 1}C_x M_{ix}. 
\]
Let us choose the coefficients as: 
\[
C_x=-\prod_{\beta\neq 1,x}\frac{(\l_1-\l_{\beta})}{(\l_x-\l_{\beta})}
\prod_{\a}\frac{(\l_x-t_{\a}-\eta)}{(\l_1-t_{\a}-\eta)}. 
\]
Notice that this expression does not depend on the index $i$.
It is necessary to calculate the following sum: 
\be
\sum_{x\neq 1}C_x M_{ix}=
-\frac{\prod_{\beta\neq 1}(\l_1-\l_{\beta})}{(t_i-\l_1+\eta)}
~\sum_{x\neq 1}f(\l_x)\prod_{\beta\neq 1,x}\left(\frac{1}{\l_x-\l_{\beta}}
\right)
\frac{1}{(t_i-\l_x)}\frac{1}{(\l_1-\l_x)}, 
\label{b1}
\ee
where
\[
f(\l_x)=
\prod_{\a\neq i}\frac{(t_{\a}-\l_x+\eta)}{(t_{\a}-\l_1+\eta)}.
\]
To calculate the sum in eq.(\ref{b1}) consider the integral in the complex 
plane over the circle of large radius which is equal to zero due to the 
behaviour of the integrand at infinity: 
\[
\oint dz\frac{f(z)}{(z-\l_1)(z-t_i)\prod_{\beta\neq 1}(z-\l_{\beta})}=0, 
~~~~~~ f(z)=\prod_{\a\neq i}\frac{(t_{\a}-z+\eta)}{(t_{\a}-\l_1+\eta)}
\]
The sum of the residues at $z=\l_{\beta}$ is equal to the sum in (\ref{b1}), 
so it is easily calculated to be 
\[
\frac{f(\l_1)}{(\l_1-t_i)\prod_{\beta\neq 1}(\l_1-\l_{\beta})} + 
\frac{f(t_i)}{(t_i-\l_1)\prod_{\beta\neq 1}(t_i-\l_{\beta})}. 
\]
Thus combining all terms we obtain the following expression for the 
new elements of the first column: 
\[
M'_{i1}=\frac{1}{(t_i-\l_1)(t_i-\l_1+\eta)}
\prod_{\beta\neq 1}\frac{(\l_1-\l_{\beta})}{(t_i-\l_{\beta})}
\prod_{\a\neq i}\frac{(t_{\a}-t_i+\eta)}
{(t_{\a}-\l_1 +\eta)}. 
\]
Note that there are several ways to choose the coefficients $C_x$ 
in order to obtain the closed expressions of this type for the first
line or column for the matrix $M$.  

 Now we are ready to come back to the sum (\ref{m}). Let us define 
$\l'_{\beta}=\l_{\beta}+\eta$ and take into account the equation 
$\det(\l_{\beta};t_k)=\det(t_k;\l'_{\beta})$. Then we get the 
following sum: 
\[
a(\l_{n_1})\sum_{\a_1}(-1)^{\a_1}\det(t_k;\l'_{\beta})
\frac{\eta}{(t_{\a_1}-\l_{n_1})(t_{\a_1}-\l_{n_1}+\eta)}
\prod\left(\frac{t_k-t_{\a_1}+\eta}{t_k-\l_{n_1}+\eta}\right)
\prod\left(\frac{\l'_{\beta}-\l_{n_1}}{\l'_{\beta}-t_{\a_1}}\right). 
\]
Comparing this sum with the last formula we find that this sum 
it equal to the determinant 
\[
a(\l_{n_1})\det(\{t\};\{\l_{n_1}\l'_{\beta}\}), 
\]
where the parameter $\l_{n_1}$ corresponds to the first column of the 
new matrix. This determinant in turn can be represented as its first 
column developement: 
\[
a(\l_{n_1})\sum_{\a_1}(-1)^{\a_1}
\frac{\eta}{(t_{\a_1}-\l_{n_1})(t_{\a_1}-\l_{n_1}+\eta)}
\det(\l_{\beta};t_k).
\]
Comparing this expression with (\ref{m}) we see that the two last 
products in (\ref{m}) can be removed - the sum over $\a_1$ does not
change. 
  Now let us come back to the formula (\ref{dd}). Instead of the 
the summation over the set $\{t_k\}$ one can take the sum over the set 
$\{t_{\a}\}$. One can label this set by the coordinates 
$\a_1,\ldots \a_{M-m}$ on the lattice consisting of $M$ sites with 
the corresponding parameters $t_1,t_2,\ldots t_M$. We can consider 
$\a_i$ as independent coordinates on the lattice, 
the only restriction being $\a_i\neq \a_j$.  The determinant 
$\det(t_{\a};\l_n)$ can be represented as the sum over the permutations 
of the products of matrix elements corresponding to the pairs of
the variables $t_{\a_i}$, $\l_{n_j}$. Each term in the sum contains 
all the variables $t_{\a_i}$ and all the variables $\l_{n_j}$. For
each term one can take the sum consequently over each of the coordinates  
$\a_1,\ldots \a_{M-m}$ with all the other indices fixed, so that 
only the set $\{k\}$ changes during the summation at each step of this 
procedure during the summation over each $\a_i$. For each $\a_i$ one 
can single out the terms with $\a_i$ and the corresponding $n_j$ from 
the last two products in the formula (\ref{dd}). Then we perform the 
procedure described above. One can see that for each term in the sum 
over the permutations (for the determinant $\det(t_{\a};\l_n)$)
this procedure removes consequently all terms in the last two products 
in the formula (\ref{dd}) and we obtain:  
\[ 
S_M(\l,t)=\frac{1}{\prod_{i<j}(t_i-t_j)}\frac{1}{\prod_{j<i}(\l_i-\l_j)}       
\sum_{k,n} (-1)^{P_k}(-1)^{P_n}  \prod(a(\l_n)) (-1)^{Mm}
\]
\be
\prod (t-\l_n+\eta) \prod (t-\l_{\beta}-\eta)
\det(\l_{\beta};t_k) \det(t_{\a};\l_n).  
\label{last}
\ee
This formula is the result of the decomposition 
of the determinant of the sum of the two matrices in (\ref{mbethe}).  
In fact, due to the symmetry of $S_M(\l,t)$ in the variables $\l_i$, 
it is sufficient to prove this statement for the set 
$\{n_0\}=(1,\ldots p), p=M-m$, and $\{\beta_0\}=(p+1,\ldots M)$. 
such that $(-1)^{P_{n_0}}=1$. Clearly, the factor $(-1)^{P_n}$ makes 
the expression for $S_M$ symmetric with respect to $\{\l\}$. 
For the term with $\{n\}=\{n_0\}$ the determinant in eq.(\ref{mbethe}) 
is represented as a sum over the permutations $P\in S_M$ with the 
sign $(-1)^P$: 
\[
S_{M}(\l,t)|_{\{n_0\}}=\frac{1}{\prod_{i<j}(t_i-t_j)\prod_{j<i}(\l_i-\l_j)} 
\prod a(\l_{n_0})f^{+}(\l_{n_0})\prod f^{-}(\l_{\beta_0})
\]
\be
(-1)^m \sum_P (-1)^P \prod a(t_{Pn_0},\l_{n_0}) 
\prod a(\l_{\beta_0},t_{P\beta_0}),
\label{mm}
\ee
where $f^{\pm}(\l_j)=\prod_{\a}(t_{\a}-\l_j\pm\eta)$ and 
$a(t_i,\l_j)$ are the matrix elements (\ref{det}). Each permutation $P$ 
can be represented as $P=P(\a)P(k)P(\{n_0\},\{\a\})$ where the permutation 
$P(\{n_0\},\{\a\})$ is 
\[
(1,2,\ldots p,p+1,\ldots M) \rightarrow 
(\a_1,\a_2,\ldots \a_p,k_1,\ldots k_m) 
\]
(here it is implied $\a_i<\a_{i+1}$, $k_i<k_{i+1}$)
and the permutations $P(\a)$, $P(k)$ corresponding to the permutations of
indices $\a_i$ and $k_i$ among themselves produce the determinants 
$\det(t_{\a};\l_{n_0})$, and $\det(\l_{\beta_0};t_k)$. The sign of the 
permutation $P(\{n_0\},\{\a\})$ differs from the sign $(-1)^{P_k}$ by 
$(-1)^{pm}$ since $(-1)^{P_k}$ corresponds to the  permutation 
$(k_1,\ldots k_m,\a_1,\ldots \a_p)$. Thus comparing the signs in the 
equations (\ref{mm}) and (\ref{last}) we see their equivalence. 
Thus we reproduce exactly the formula (\ref{mbethe}).

\vspace{0.1in}

{\bf Acknowledgements }

The author is deeply indebted to the stuff of INR Theory Division   
for the support.

\end{document}